\documentstyle[12pt]{article}
\begin{document}
\pagestyle{empty}
\begin{center}
\hspace {10.cm} DFUPG 99-95
\vspace {2 cm.}
{\large\bf Polyakov Loops in 2D QCD}\\
\vspace {1 cm}
{\large G. Grignani\footnote{Permanent address: Dipartimento di Fisica
and Sezione I.N.F.N., Universit\'a di Perugia, Via A. Pascoli I-06100
Perugia, Italy}, G. Semenoff and P. Sodano$^1$}\\
\vspace {0.4cm}
Department of Physics,
University of British Columbia\\Vancouver, British Columbia, Canada V6T 1Z1
\\
\vspace {0.4cm}
\end{center}
\vspace {1 cm}
\centerline{\bf Abstract}
\vspace {0.4 cm}

We discuss SU(N) gluo-dynamics at finite temperature and on a spatial
circle.  We show that the effective action for the Polyakov Loop
operator is a one dimensional gauged SU(N) principle chiral model with
variables in the loop space and loop algebra of the gauge group. We
find that the quantum states can be characterized by a discrete
$\theta$-angle which appears with a particular 1-dimensional
topological term in the effective action.  We present an explicit
computation of the partition function and obtain the spectrum of the
model, together with its dependence on the discrete theta angle.  We
also present explicit formulae for 2-point correlators of Polyakov
loop operators and an algorithm for computing all N-point correlators.

\newpage
Yang-Mills theory in 1+1-dimensions is a prototypical example of a
topological field theory.  The classical theory is trivial and the
quantum theory has no propagating degrees of freedom.  It nevertheless
has interesting features, particularly in the large $N$ limit or when
it is defined on a multiply connected spacetime ~\cite{w,bt} where its
partition function depends on the area as well as topological
invariants.  On a sphere it exhibits a large $N$ phase transition for
some critical value of $eA$ where $e$ is the coupling and $A$ is the
area of the sphere ~\cite{dk}.  Wilson loops can be computed in both
the strong and weak coupling phases and exhibit interesting behaviors
{}~\cite{b,dk1}.  It has been shown that the strong coupling expansion
can be rewritten as a lower-dimensional string theory ~\cite{gt}.
Furthermore, when defined on Riemann surfaces with non-zero genus it
is known to have degrees of freedom related to the gauge group
holonomy on the homology cycles of the surface ~\cite{w,bt}.  In
particular, on cylindrical spacetime, it can be solved explicitly and
has quantum mechanical degrees of freedom corresponding to the
eigenvalues of the Wilson loop operators which wind around the compact
spatial direction ~\cite{ra}-\cite{da}.  In that case, topology also
affects the gauge fixing, ~\cite{ls}.  Gauge invariant two dimensional
models are also intimately connected with one dimensional integrable
quantum systems ~\cite{pol,ap}.

In this paper we shall examine Yang-Mills theory on a torus.  It
corresponds to finite temperature 1+1-dimensional gluo-dynamics on a
spatial circle.  We show that, in any 1+1-dimensional gauge theory,
correlators of Polyakov loop operators, which are the Wilson loop
operators that wind around the periodic time, can be
computed as correlators in a particular 1-dimensional gauged principle
chiral model.  In either pure Yang-Mills theory, or in Yang-Mills
theory coupled to matter which is in the adjoint representation of the
gauge group, the effective field theory has a global symmetry which
transforms the Polyakov loop operator by elements of the center of the
gauge group.  For the case of pure Yang-Mills theory, we find the
effective field theory explicitly.  We also find a solution of the
effective theory and use it to give an explicit computation of the
partition function and the two-point correlators of Polyakov loop
operators as well as an algorithm for computing all correlators of
Polyakov loop operators.

A few years ago, Polyakov ~\cite{1.} and Susskind ~\cite{5.} suggested
that the question of confinement in quantum gluo-dynamics at finite
temperature is intimately related to the realization of a global
symmetry involving the center of the gauge group.  This symmetry
governs the expectation value of the Polyakov loop operator
\begin{equation}
P(x)={\rm tr}{\cal P}\exp\left(i\int_0^{1/T}d\tau A_0(x,\tau)\right)
\label{l}
\end{equation}
which measures the gauge group holonomy in the periodic Matsubara
time,
\begin{eqnarray}
<P(x_1)\ldots P(x_m)P^{\dagger}(y_1)\ldots P^{\dagger}(y_n)>=
{}~~~~~~~~~~~~~~~~~~~~~~~~~~~~~~~~~~~~~~~~~~~~~\nonumber
\\={\int
dA_\mu e^{-\int_0^{1/T} d\tau dy {\rm tr}F_{\mu\nu}^2(\tau,y)/2}
P(x_1)\ldots P(x_m)P^{\dagger}(y_1)\ldots P^{\dagger}(y_n)
\over\int dA_\mu e^{-\int_0^{1/T} d\tau dy {\rm tr}F_{\mu\nu}^2(\tau,y)/2} }
\label{defn}
\end{eqnarray}
(The Euclidean path integral has periodic boundary conditions,
$A_\mu(1/T, \vec x)=A_\mu(0,\vec x)$ and $T$ is the temperature.  For
a discussion of the path integral formulation of finite-temperature
gauge theory, see ~\cite{2.}.)  The action and measure in the path
integral (\ref{defn}) are symmetric under gauge transformations
\begin{equation}
A_\mu(\tau,x)\rightarrow g^{\dagger}(\tau,x)A_\mu(\tau,
x)g(\tau,x)+ig^{\dagger}(\tau,x)
\nabla_\mu g(\tau,x)
\end{equation}
where $g(\tau,x) $ is a gauge group element (here we consider SU(N))
which is periodic up to an element of the center of the gauge group,
$Z(N)$,
\begin{equation}
g({1/T},x)=g(0,x)e^{2\pi i n/N}
\end{equation}
The path-ordered exponential transforms as
\begin{equation}
{\rm tr}{\cal P} e^{i\int_0^{1/T} d\tau A_0( x,\tau)}\rightarrow {\rm
tr}g^{\dagger}({1/T}, x){\cal P} e^{i\int_0^{1/T} d\tau A_0(
x,\tau)}g(0, x) =e^{2\pi in/N}{\rm tr}{\cal P} e^{i\int_0^{1/T} d\tau
A_0( x,\tau)}
\end{equation}
The expectation value (\ref{defn}) is interpreted as the free energy
of the system in the presence of an array of external fundamental
representation quark sources at positions $x_1,\ldots,x_m$ and
anti-quarks at $y_1,\ldots,y_n$.  The realization of the $Z(N)$
symmetry is related to confinement - it is represented faithfully in a
confined phase and it is spontaneously broken in a de-confined phase.
In 3+1-dimensional chromodynamics a first order phase transition
between confining and deconfining phases is expected at some critical
temperature.  The domain walls and Z(N) bubbles between different
$Z_N$ vacua that should appear in the de-confined, broken $Z(N)$
phase, have been extensively investigated ~\cite{3.,4.}.

On the other hand, 1+1-dimensional gluo-dynamics is confining and the
global $Z(N)$ symmetry is unbroken at all temperatures.  There, the
correlators of Polyakov loop operators are known to be related to the
correlators in a generalized Sutherland model ~\cite{pol}.

The Hamiltonian of 1+1-dimensional gluo-dynamics is
\begin{equation}
H=\frac{e^2}{2} \int_0^L dx \sum_a E^a(x)^2
\label{ham}
\end{equation}
with nonvanishing commutator
\begin{equation}
[ A^a(x),E^b(y)]=i\delta^{ab}\delta(x-y)
\label{comm}
\end{equation}
gauge constraint
\begin{equation}
{\cal G}^a(x)\equiv\nabla E^a(x)+f^{abc}A^b(x)E^c(x)\sim0
\label{gauge}
\end{equation}
and periodic boundary conditions, $E^a(L)=E^a(0)$, $A^a(L)=A^a(0)$.
In Eqs.(\ref{ham},\ref{comm},\ref{gauge}) $a,b=1\ldots N^2-1$, the
traceless hermitean generators $T^a$ are normalized so that ${\rm
tr}T^aT^b=\frac{1}{2}\delta^{ab}$, the structure constants are defined
by $[ T^a,T^b]=if^{abc}T^c$ and we define the $N\times N$ matrices
$E(x)\equiv E^a(x)T^a$, $A(x)\equiv A^a(x)T^a$.

Since $A(x)$ and $E(x)$ transform under the adjoint action of the
gauge group, their periodic boundary conditions are preserved by time
independent gauge transformations which are periodic up to an element
of the center of the group, $g_n(L)= g_n(0)e^{2\pi i n/N}$.
$G_\chi=\int_0^Ldx G^a(x)\chi^a(x)$ commutes with the Hamiltonian and
generates time independent, periodic gauge transforms
\begin{eqnarray}
e^{iG_\chi}E(x)e^{-iG_\chi}= g_0^{\dagger}(x)E(x)g_0(x) \\ e^{iG_\chi}
A(x)e^{-iG_\chi}= g_0^{\dagger}(x)A(x)g_0(x)+ig_0^{\dagger}(x)\nabla
g_0(x)
\end{eqnarray}
where $g_0(x)=1-i\chi(x)+\ldots$ and where $g_0(0)=g_0(L)$.  The
constraint (\ref{gauge}) indicates that the physical states are
invariant under all strictly periodic gauge transforms.  The coset
group of all gauge transformations modulo periodic ones is isomorphic
to the center of the gauge group, $Z_N$.  The physical states must
transform under an irreducible unitary representation of the coset.
Representations of $Z_N$ are one dimensional and are characterized by
a discrete angle.\footnote{The number of connected components of the
group of periodic gauge transformations, which is the loop group based
on $SU(N)/Z_N$, is
\begin{equation}
\Pi_0( {\rm loops~on}~SU(N)/Z_N)=\Pi_1(SU(N)/Z_N)=\Pi_0(Z_N)=Z_N
\end{equation}
The group $Z_N$ has $N$ 1-dimensional irreducible representations,
labelled by an integer $m=0,1,\ldots,N-1$ and where the elements are
$U(n;\theta_m)=e^{i\theta_m n}$ and $\theta_m=2\pi m/N$.}

This theta-angle will also characterize the quantization of Yang-Mills
theory with coupling to any adjoint representation matter fields.
Some related issues have been pointed out in ~\cite{sh}.  Also, for
adjoint QCD on a hypertorus in d spatial dimensions, there are d
angles related to periodicity of gauge transformations as well as
perhaps others related to higher dimensional winding numbers.  In
either case, if one introduces matter fields which are in the
fundamental representation, their boundary conditions reduce the gauge
symmetry to gauge transformations with a fixed periodicity, i.e. one
of the degenerate states, and there is no theta-angle.

In the Schr\"odinger picture, where states are wavefunctionals
$\Psi[A]$, and the electric field is realized as $E^a(x)\equiv
\frac{1}{i}\frac{\delta} {\delta A^a(x)}$, the physical state
condition is
\begin{equation}
\left(\frac{1}{i}\nabla \frac{\delta}{\delta A^a(x)} +f^{abc}A^b(x)
\frac{\delta}{\delta A^c(x)}\right) ~\Psi_{\rm phys}[A]=0
\end{equation}
The wavefunctionals of physical states therefore transform as
\begin{equation}
\Psi_{\rm phys}[ g_n^{\dagger}Ag_n+ig_n^{\dagger}\nabla g_n;\theta]=
e^{in\theta}\Psi[A;\theta]
\end{equation}

The thermodynamic partition function is obtained by taking a trace of
the Botlzmann factor, $e^{-H/T}$, over the physical states.  This is
most conveniently expressed as a trace using eigenstates of $A(x)$ and
a projection onto gauge invariant theta-states,
\begin{equation}
Z= \frac{1}{{\rm VOL~ G}}\sum_n\int [dg_n]e^{-i\theta n}~\int\prod_x
dA^a(x)~<A\vert e^{-H/T}
\vert g^{\dagger}_nAg_n+ig_n^{\dagger}\nabla g_n>
\label{part}
\end{equation}
The integration over gauge transforms projects onto physical states
with angle $\theta$.  This integral can be expressed in terms of the
standard Euclidean path integral (see ~\cite{3.}) which is used to
take expectation values of observables as in (\ref{defn}) by first
considering the integral with twisted boundary conditions,
\begin{equation}
Z=\frac{1}{\rm VOL~G}\sum_n\int[dg_n]e^{-i\theta n}\prod_{x,\tau}
dA^a(x,\tau) e^{-\int_0^{1/T}d\tau dx\frac{1}{e^2}{\rm tr}{\dot
A}^2(x,\tau)}
\label{pi}
\end{equation}
\begin{equation}
A(1/T,x)=g_n^{\dagger}(x)A(0,x)g_n(x)+ig_n^{\dagger}(x)\nabla g_n(x)
\end{equation}
The boundary conditions can be made periodic by performing a
non-periodic gauge transformation.
\begin{equation}
A(\tau,x)\rightarrow
g_n^{\dagger}(\tau,x)A(\tau,x)g_n(\tau,x)+
ig_n^{\dagger}(\tau,x)\nabla g_n(\tau,x)
\end{equation}
where $g_n(1/T,x)=g_n(0,x)g_n(x)$.  The result is the standard form of
the path integral in (\ref{defn}).
The loop operator is identified as
\begin{equation}
{\cal P}e^{i\int_0^{1/T}d\tau A_0(\tau,x)}
\equiv g_n(x)
\end{equation}
and
the trace of the Polyakov loop operator is equal to the
trace of the gauge group valued variable in (\ref{part}).
\begin{equation}
P(x)={\rm tr} g_n(x)
\end{equation}

The integrand in (\ref{part}),
\begin{equation}
e^{-i\theta n}<A\vert e^{-H/T}\vert g_n^{\dagger}Ag_n+ig_n^{\dagger}\nabla
g_n>\equiv
e^{ -S_{\rm eff}[A,g_n]} \frac{1}{N}{\rm tr}\left(
g_n(L)g_n^{\dagger}(0)\right)^{-\theta N/2\pi}
\label{defs}
\end{equation}
is a functional of $A$ and $g_n$ which has the symmetry properties
\begin{equation}
S_{\rm eff}[A,g_n]= S_{\rm eff}[u^{\dagger}Au+iu^{\dagger}\nabla u,u^{-1}g_nu]
\label{s1}
\end{equation}
\begin{equation}
S_{\rm eff}[A,g_nz]=S_{\rm eff}[A,g_n]
\label{s2}
\end{equation}
The variables $A$ and $g_n$ take values in the loop algebra and loop
group of the gauge group, respectively.  This is a 1-dimensional
gauged principle chiral model.
The theta-term in (\ref{defs}) is a topological term in the
sense that it is insensitive to local variations of $g_n(x)$
and depends only on its holonomy on the circle. (\ref{s2})
implies that the effective
action is invariant under the $Z_N$ symmetry $g_n\rightarrow g_nz$.  The
expectation values of Polyakov loop operators are given by the
correlators of traces of the group elements
\begin{eqnarray}
<P(x_1)\ldots P^{\dagger}(y_k)>=~~~~~~~~~~~~~~~~~~~~~~~~~~~~~~~~~~~~~~~~~
\nonumber\\
\frac{\sum_n \int dA(x)[dg_n(x)]e^{-S_{\rm eff}[A,g_n]}\frac{1}{N}{\rm tr}
\left(g_n(L)g_n^{\dagger}
(0)\right)^{-\theta N/2\pi} {\rm tr}g_n(x_1)\ldots {\rm tr}g_n^{\dagger}(y_k)
}{
\sum_n\int dA(x)[dg_n(x)]e^{-S_{\rm eff}[A,g_n]}
\frac{1}{N}{\rm tr}\left(g_n(L)g_n^{\dagger}(0)\right)^{-\theta N/2\pi}  }
\end{eqnarray}

The equation (\ref{defs}) defines an d-dimensional effective action
for the Polyakov loop operator in {\it any} d+1-dimensional gauge theory
coupled to matter fields where there is a trace over the matter
degrees of freedom on the right hand side of the formula.  In all cases, the
resulting effective action has the gauge invariance property in (\ref{s1}).
When the matter is in the adjoint representation, or any other representation
which is invariant under transformations in the center of the
gauge group, the effective action also has the global
$Z_N$ symmetry in (\ref{s2}).
In the present paper, we shall concentrate on finding the effective action
for pure 1+1-dimensional Yang-Mills theory.

By evaluating the matrix element in (\ref{part}),
\begin{eqnarray}
<A\vert \exp\left(\frac{-e^2}{2T}\sum_a\int_0^Ldx
\frac{-\delta^2}{\delta A^a(x)^2}
\right)
\vert g_n^{\dagger}Ag_n+ig_n^{\dagger}\nabla g_n>\\=\exp
\left( -\frac{T}{e^2}\int_0^L dx~{\rm tr}(A-g_n^{\dagger}Ag_n-
ig_n^{\dagger}\nabla g_n)^2\right)
\end{eqnarray}
(when defined using zeta function regularization, the constant factor
which should appear on the right-hand-side is one) we present the
partition function as
\begin{eqnarray}
Z[\theta,T,L]= \frac{1}{{\rm VOL~ G}}\sum_{n}\int \prod_{x,a}
dA^a(x)[dg_n(x)]~~~~~~~~~~~~~~~~~~~~~~~~~~~~~~~\nonumber\\
\exp\left( -\frac{T}{e^2}\int_0^L
dx~{\rm tr}\vert Dg_n\vert^2\right){\rm tr}\left(g_n(L)g_n^{\dagger}(0)
\right)^{-\theta N/2\pi}
\label{prin}
\end{eqnarray}
where $Dg_n=\nabla g_n +[A,g_n]$.

The integral in (\ref{prin}) can be done by using gauge symmetry to
diagonalize the unitary matrix $g_n=Ug_n^DU^{-1}$ where $g_n^D(x)={\rm
diagonal}(e^{i\phi_n^1(x)},
e^{i\phi_n^2(x)},\ldots,e^{i\phi_n^N(x)})$.  The $\phi$ variables are
angles $\phi\in[0,2\pi]$.  The measure in the integral has the form
\begin{equation}
[dg_n(x)]=\prod_{x\in[0,L]}\prod_\alpha
d\phi_n^\alpha(x)\prod_{\alpha<\beta}\frac{1}{2}
\sin^2(\phi_n^\alpha(x)-\phi_n^\beta(x))
\delta(\sum_\alpha\phi_n^\alpha)
[dU(x)]\ \ ,
\end{equation}
and the action is $\frac{T}{e^2}\int_0^Ldx
\left(\sum_\alpha \nabla
\phi_n^\alpha\nabla\phi_n^\alpha+\sum_{\alpha,\beta}\vert
A_{\alpha
\beta}\vert^2\vert
e^{i\phi_n^\alpha}-e^{i\phi_n^\beta}\vert^2\right)$.  The integral
over $A_{\alpha\beta}$, where $\alpha\neq\beta$, cancels the
Vandermonde determinant in the integration measure.  The integration
over diagonal components of $A$ yields an infinite factor which
compensates the infinite normalization of the plane-wave state which
was used to take the trace in (\ref{part}).  The partition function is
an integral over the eigenvalue variables,
\begin{equation}
Z=\frac{1}{\rm VOL ~G} \sum_{n=0}^{N-1} e^{-i\theta n}\int
\prod_{x}\prod_{\alpha=1}^{N-1} d\phi_n^\alpha(x)
\exp\left[-T/e^2\int_0^L dx \sum_{\alpha=1}^{N-1}\nabla
\phi_n^\alpha\left(\nabla\phi_n^\alpha+
\sum_{\beta=1}^{N-1}\nabla\phi_n^\beta\right)\right]
\label{zsun}
\end{equation}
where the integration variables have the boundary condition
$\phi_n^\alpha(L)= \phi_n^\alpha(0)+ \theta n + 2\pi p_\alpha$ ,
$\theta=2\pi m/N$ and $p\in Z\hspace{-2mm}Z$ and the integration
measure integrates $\phi(x)$, at each point x, over the range
$[0,2\pi)$ and sums over the integers $p_\alpha$.  The integral is
invariant under the field translation symmetry
\begin{equation}
\phi_n^\alpha(x)
\rightarrow \phi_n^\alpha(x)+c^\alpha
\end{equation}
where $c^\alpha$ is a constant.  Note that this is not a symmetry of
the action or the integration measure in the path integral
(\ref{prin}) separately but only appears after the integration over
$A$.  This symmetry can be used to extend the limits on the
integration over $\phi(x)$ at each $x$ to infinity.  One can untwist
the boundary conditions and take into account the periodicity of the
$\phi^\alpha_n(x)$ variables, by changing the functional integration
variables to $\phi^\alpha(x)$, by means of
\begin{equation}
\phi_n^\alpha(x)=\frac{2\pi (n+Np_\alpha) x}{NL} +\phi^\alpha(x)\ \ ,
\label{boun}
\end{equation}
with $p_\alpha\in Z\hspace{-2mm}Z$. In (\ref{boun}) $\phi^\alpha(x)$
is real and periodic, $\phi^\alpha(L)=\phi^\alpha(0)$.  The partition
function can be written as the product of an analytic part $Z_a$
containing the integral over the periodic fields $\phi(x)$ and a
topological part $Z_t$ containing the summation over integers
$p_\alpha$, $Z=Z_a Z_t$.  $Z_a$ can be computed by means of a zeta
function regularization.  Expanding $\phi^\alpha(x)$ in modes
\begin{equation}
\phi^\alpha(x)={\frac{1}{\sqrt{L}}}\sum_{k=-\infty}^\infty
a_k e^{i2\pi k x/L} \,
\end{equation}
the functional measure is defined as $\prod_xd\phi(x)\equiv
\prod_k da_k$.  The integration over the zero mode produces an
infinite, irrelevant, temperature independent factor proportional to
the volume of the gauge group.  The functional integral in $Z_a$ is
proportional to the determinant of the laplacian,
\begin{equation}
Z_a=\prod_{k\neq 0}\left(\det \frac{N^2}{i}\Omega k^2\right)^{-\frac{1}{2}}
=\det\left( \frac{N^2}{i}\Omega\right)^{-\zeta(0)}e^{(N-1)\zeta'(0)}=
\sqrt{N}\left(\frac{2T}{e^2 L}\right)^{\frac{N-1}{2}}\ \ .
\end{equation}
where $\Omega$ is the $(N-1)\times(N-1)$ matrix
\begin{equation}
\Omega=i\frac{4\pi T}{e^2 L N^2}\pmatrix{2&1&\ldots&1\cr
1&2&\ldots&1\cr
\vdots&\vdots&\ddots&\vdots\cr
1&1&\ldots&2\cr}\equiv ig_{ij} \ \ .
\label{omega}
\end{equation}
and we have used zeta-function regularization with
\begin{equation}
\zeta(s)=\sum_{k=1}^{\infty}\frac{1}{k^{2s}}
\end{equation}
and $\zeta(0)=-1/2$ and $\zeta'(0)=-(\log 2\pi)/2$.

$Z_t$ is given by a finite sum over $n$ and an infinite sum over all
the $p_\alpha$. Rewriting
\begin{equation}
\exp\left(-\frac{i 2\pi m n}{N}\right)=\exp\left[i 2\pi
m\sum_{\alpha=1}^{N-1}(n+Np_\alpha)/N\right]\ \ ,
\end{equation}
the finite sum over $n$ and the $N-1$ infinite sums over the
$p_\alpha$ become just $N-1$ infinite sums over the integers
$n_\alpha=n+Np_\alpha$. The $n_\alpha$ in fact span $Z\hspace{-2mm}Z$
when $0\le n\le N-1$ and $p\in Z\hspace{-2mm}Z$.  $Z_t$ then reads
\begin{equation}
Z_t=\sum_{\vec n=-\infty}^{+\infty}\exp\left(2\pi i \vec n^t\cdot\vec
z+\pi i
\vec n^t\Omega\vec n \right)\equiv\theta\left(\vec z,\Omega\right)\ \ .
\end{equation}
Here we introduced the $\theta$-function of several variables
according to the conventions of Ref.~\cite{7.}, $\vec n$, $\vec z$ are
thought of as $N-1$-dimensional column vectors, $\vec
n^t=(n_1,n_2,\cdots,n_{N-1})$, $\vec z^t=(m/N,m/N,\cdots,m/N)$.

In order to obtain the energy spectrum of the theory we can now apply
the generalized Poisson resummation formula
\begin{equation}
\sum_{\vec m=-\infty}^{+\infty}\exp\left(-\pi g_{ij}m^i m^j -2\pi i
m^ia_i\right)=\frac{1}{\sqrt{\det g_{ij}}}
\sum_{\vec n=-\infty}^{+\infty}\exp\left[-\pi
g^{ij}(n_i-a_i)(n_j-a_j)\right]. \label{pois}
\end{equation}
where $g^{ij}$ is the inverse of $g_{ij}$
\begin{equation}
g^{ij}=\frac{e^2 L N}{4\pi T}\pmatrix{N-1&-1&\ldots&-1\cr
-1&N-1&\ldots&-1\cr
\vdots&\vdots&\ddots&\vdots\cr
-1&-1&\ldots&N-1\cr} \ \ .
\label{ginv}
\end{equation}
and $\det g_{ij}=(4\pi T/e^2l N^2)^{(N-1)} N$.  The $\sqrt{\det
g_{ij}}$ cancels the temperature dependence of $Z_a$ so that the
partition function reads
\begin{equation}
Z=\sum_{\vec n=-\infty}^{+\infty}\exp\left\{-\frac{e^2
L}{4T}\left[\sum_{\alpha=1}^{N-1}(Nn_\alpha+m)^2-\frac{1}{N}
\left(\sum_{\alpha=1}^{N-1}(Nn_\alpha+m)\right)^2\right]\right\}\ \ .
\label{zeta}
\end{equation}
{}From (\ref{zeta}) one can deduce the energy spectrum of the theory for
each $m$-sector. The result is energy levels which are characterized
by $N-1$ integers $\vec n$ and
\begin{equation}
E(\vec n)=\frac{e^2L}{4}\left[
\sum_{\alpha=1}^{N-1}\left(Nn_\alpha+m\right)^2-\frac{1}{N}\left(
\sum_{\alpha=1}^{N-1}\left(Nn_\alpha+m\right)\right)^2\right]
\end{equation}
which agrees with the one obtained, in a different context, in
Refs.~\cite{ra,hh,he} with the small difference that one finds that
each energy eigenvalue belongs to a specific topological sector,
labelled by $m$.  For example for $SU(2)$ in Refs.~\cite{ra,hh,he}
was found $E_n=e^2 Ln^2/8$ and here we see that the eigenvalues
corresponding to even integers belong to the $m=0$ sector ($E_n=e^2 L
(2n)^2/8$) those corresponding to odd integers ($E_n=e^2 L
(2n+1)^2/8$) belong to the $m=1$ sector.

The calculation of correlators of Polyakov loop operators
straightforward.  The path integral is easily performed and the sums
on $n$ and the $p_\alpha$ can be treated as before.

Correlators are nonvanishing only if they are invariant under the
transformations by the center of the group, $Z(N)$, in (\ref{s2}).
This occurs when the number of $g$'s in a correlator is equal to the
number of $g^{\dagger}$'s, modulo $N$.  Using Eqs.(\ref{pois}), the
2-point correlator $P^{(2)}_m(x,y)=<tr\{g(x)\}tr\{g^{\dagger}(y)\}>_m/N^2$
can be presented as
\begin{eqnarray}
P_m^{(2)}(x,y)=\frac{1}{ZN^2}\sum_{\vec n=-\infty}^{+\infty}\left\{
\exp\left[-\pi g^{ij}\left(n_i+\frac{m}{N}
-\frac{x-y}{NL}\right)\left(n_j+\frac{m}{N}
-\frac{x-y}{NL}\right)\right]\right.\nonumber\\+
\left.\sum_{\alpha=1}^{N-1}\exp\left[-\pi g^{ij}\left(n_i+\frac{m}{N}
-\frac{x-y}{NL}\delta_{i\alpha}\right)\left(n_j+\frac{m}{N}
-\frac{x-y}{NL}\delta_{j\alpha}\right)\right]\right\}G(x,y)\ \ ,
\end{eqnarray}
where
\begin{equation}
G(x,y)=\exp\left[-\frac{L
e^2(N-1)}{4TN}\left(\frac{|x-y|}{L}-\frac{(x-y)^2}{L^2}\right)\right]\ \ .
\end{equation}
For $SU(2)$
\begin{equation}
P_m^{(2)}(x,y)=
\frac{1}{2Z}\sum_{n=-\infty}^{+\infty}
\exp\left\{-\frac{e^2L}{8T}\left[(2n+m)^2-
(4n+2m)\left(\frac{x-y}{L}\right)+\frac{|x-y|}{L}\right]\right\}\ .
\end{equation}
Any other correlator of  Polyakov loop operators can be analogously
obtained.

P. Sodano and G. Grignani acknowledge the hospitality of the Physics
Department of the University of British Columbia where part of this
work was done. G. Semenoff acknowledges the hospitality of the
Dipartimento di Fisica of the Universit\`a di Perugia.  The authors also
acknowledge financial support of NATO, NSERC and the INFN.


\begin{thebibliography}{}
\bibitem{w}E. Witten, Comm. Math. Phys. 141, 153, 1991; Jour.
Geom. Phys. 9, 303, 1992.
\bibitem{bt}M. Blau and G. Thompson, Proceedings of Summer School in High
Energy Physics and Cohomology, Trieste, Italy, 1993 (hepth/9310144).
\bibitem{dk}M. Douglas and V. Kazakov, Phys. Lett. 319B, 1993, 219.
\bibitem{b}D. Boulatov, Mod. Phys. Lett. A9, 365, 1994.
\bibitem{dk1}J.-M. Daul and V. Kazakov, preprint LPTENS-9337, hepth/9310165.
\bibitem{gt}G. Gross and W. Taylor, Nucl. Phys. B400, 181, 1993;
Nucl. Phys. B403, 395, 1993.
\bibitem{ra}S. G. Rajeev, Phys. Lett. B212, 203 (1988);
\bibitem{hh}J. E. Hetrick and Y. Hosotani, Phys. Lett. B230, 88 (1989).
\bibitem{he}J. E. Hetrick, Int. J. Mod. Phys. A9, 3153 (1994).
\bibitem{da}M. Caselle, A. D'Adda, L. Magnea and S. Panzeri,
Torino Preprint DFTT-50-93 (hepth/9309107).
\bibitem{ls}E. Langmann and G. Semenoff, Phys. Lett. B296, 117 (1992);
Phys. Lett. B303, 303 (1993).
\bibitem{pol}J. Minahan and A. Polychronakos, Phys. Lett. B212, 155 (1993);
Nucl. Phys. B422, 172 (1994).
\bibitem{ap}A. Niemi and P. Pasanen, Phys. Lett. B323, 46 (1994).
\bibitem{1.}A.M. Polyakov, Phys. Lett. 72B, 477 (1978).
\bibitem{5.}L. Susskind, Phys. Rev. D20, 2610 (1979).
\bibitem{2.}D. J. Gross, R. D. Pisarski and L. G. Yaffe, Rev. Mod.
Phys. 53, 43 (1981).
\bibitem{3.}T. Bhattacharya, A. Gocksch, C. Korthals-Altes and R.
D. Pisarski, Phys. Rev. Lett. 66, 998 (1991).
\bibitem{4.}V. Dixit and M. G. Ogilvie, Phys. Lett. B269, 353 (1991);
J. Ignatius, K. Kajantie and K. Rummukainen, Phys. Rev. Lett.  68, 737
(1992).
\bibitem{sh}F. Lenz, M. Shifman and M. Thies, UPI-MINN-94/34-T,
UMN-TH-1315-94, CTP-2391, 1994.
\bibitem{7.}D. Mumford,{\it Tata Lectures on Theta}, Birkh\"auser,
Boston, 1983.
 \end{thebibliography}
\end{document}